\documentclass[11pt,fleqn]{article}
\title{Coherent Causal Memory}
\author{Ernie Cohen\\University of Pennsylvania\\(ernie.cohen@acm.org)}
\usepackage{ifthen}
\usepackage{comment}
\usepackage{abbrev}
\usepackage{latexsym}
\usepackage{amssymb}
\usepackage[margin=1.3in]{geometry}
\newcommand\Def[1]{{\em #1}}
\newcommand\Ref[1]{(\ref{#1})}
\newcommand\False{\mathit{false}}
\newcommand\True{\mathit{true}}

\newcommand\AndSym{\land}
\newcommand\OrSym{\lor}

\newcommand\ImpliesSym{\Rightarrow}
\newcommand\ImpliedSym{\Leftarrow}
\newcommand\IffSym{\Leftrightarrow}

\newcommand\Induc{\mbox{induc hyp}}

\newenvironment{Proof}{
\[\begin{array}{l@{\hspace{5mm}}l@{\hspace{5mm}\{}l@{\}}}}{\end{array}\]}

\newcommand\Set[1]{\{#1\}}

\begin{abbreviate}{Math}{~(=<>/}{~(=<>/}
  \x => {\ImpliesSym}
  \x == {\IffSym}
  \x =   {=}
  \x <=>  {\IffSym}
  \x <=  {\ImpliedSym}
  \x <>  {\Diamond}
  \x <x> {\X}
  \x <x  {<x}
  \x <   {<}
  \x >   {>}
  \x /|  {\wedge}
  \x /   {/}
  \x ~   {\neg}
  \x (E: {(\exists}
  \x (E  {(E}
  \x (A: {(\forall}
  \x (A  {(A}
  \x (Q: {(\Kw{Q}}
  \x (Q  {(Q}
  \x (   {(}
\end{abbreviate}
\Math

\def\Kw#1{\mbox{{\bf #1}}}
\def\/{\vee}
\def\AndSym{\wedge}
\def\OrSym{\vee}
\def\ImpliesSym{\Rightarrow}
\def\ImpliedSym{\Leftarrow}
\def\IffSym{\Leftrightarrow}

\def\Ref#1{(\ref{#1})}
\def\Def#1{{\em #1}}

\def\Red#1{\langle #1 \rangle}

\def\Prestate#1#2#3{\mathrm{prestate}(P,o,s)}

\def\True{\mathit{true}}
\def\False{\mathit{false}}

\def\Wait#1{{\bf \mathbf{wait}}(#1)}
\def\Skip{}

\def\Rem#1#2{#1-#2}

\def\Op#1#2{#1 \rightarrow #2}
\def\B:#1:#2{{#1}_{#2}}
\def\NA:#1:#2{{#1}^{#2}}
\def\Assume#1{\mathit{assume}(#1)}
\def\At#1#2{\mathit{at}(#1,#2)}
\def\SB#1{\langle #1 \rangle}
\begin{document}
\maketitle
\begin{abstract}
Coherent causal memory (CCM) is causal memory in which prefixes of an execution can be mapped to global memory states in a consistent way. While CCM requires conflicting pairs of writes to be globally ordered, it allows writes to remain unordered with respect to both reads and nonconflicting writes. Nevertheless, it supports assertional, state-based program reasoning using generalized Owicki-Gries proof outlines (where assertions can be attached to any causal program edge). Indeed, we show that from a reasoning standpoint, CCM differs from sequentially consistent (SC) memory only in that ghost code added by the user is not allowed to introduce new write-write races.

While CCM provides most of the formal reasoning leverage of SC memory, it is much more efficiently implemented.   As an illustration, we describe a simple programming discipline that provides CCM on top of x86-TSO. The discipline is considerably more relaxed than the one needed to ensure SC; for example, it introduces no burden whatsoever for programs in which at most one thread writes to any variable.
\end{abstract}
\Math

\section{Introduction}

Consider the following simple multithreaded program, started from a state where $~x /| ~y$:
\begin{eqnarray*}
&&\mathbf{cobegin}\\
&&\ \ \  \; x := \True;  \Wait{\neg y}; \\
&&\ ||\ y := \True;  \Wait{\neg x};\\
&&\mathbf{coend}
\end{eqnarray*}
On sequentially consistent (SC) memory (SCM) \cite{SC}, where
all operations from all threads are linearly ordered, this program
has no complete executions; whichever thread performs its assignment second will get stuck
waiting forever. However, in weaker memory models such as TSO (by which we mean "total store
order with store buffer forwarding", the memory model for processors in the x86 family \cite{x86TSO}),
store buffering can introduce a delay in the write of a thread being seen by another thread, 
so there is an execution in which both writes happen (their stores
entering their respective store buffers), then both reads (i.e. waits) complete (since neither store has yet reached
the shared memory), before the stores hit memory.
Similar behaviors are possible in other weak memory models, as well as distributed memory models
that expose the latency between assignment to a variable in one node and its appearance as an update in another node.

Now, suppose that instead of thinking operationally about such programs, we want to reason about them using ordinary 
state assertions.
The usual way to reason about a concurrent or distributed algorithm is to just give a big global invariant,
one that takes into account the program counters and so on. But it is easy to see that such reasoning is not
only sound for SCM, it is also \Def{complete}, and so cannot be used to reason about weaker memory models.
(The completeness immediately follows from the invariance of "the current state is reachable from the initial state via a sequential
execution".) 

Because invariance reasoning is rather fundamental, it is usually taken for granted that
the right way to deal with weak memory is to get back to the world of SC. One way to do this is to make the underlying weak
memory explicit (e.g. by exposing the store buffers as part of the program state), a solution that makes reasoning
even about trivial programs like this one a painful experience. Another way is to require programs to obey a discipline
that guarantees SC (e.g., \cite{SBReduction}); in the example above, such a discipline would require each of the threads
to flush their store buffers between their writes to shared memory and their subsequent reads. 
This solves the reasoning problem, but at the cost of possibly slowing down code unnecessarily.

So let's return to reasoning about this program. What makes weak memory weak is the semantic importance
attached to thread boundaries, something we lose when we reason with global invariants. 
To reason in a weak memory model, we need a more thread-centric approach to reasoning that treats
actions within a thread differently from actions from different threads.
One of the earliest such approaches, due to Owicki and Gries (OG) \cite{OG}, is to reason about each thread 
as if it was a sequential program, but to require an additional check that actions of other threads don't break any of the
intermediate assertions. For example, we could annotate the program above as follows:

\begin{eqnarray*}
&& \Set{~x /| ~y}\\
&&\mathbf{cobegin} \\
&&\ \ \  \; \Set{~x}\  x := \True;  \Set{x}\ \Wait{~y};  \Set{x}\\
&&\ ||\ \Set{~y}\ y := \True;  \Set{y}\ \Wait{~x}; \Set{y}\\
&&\mathbf{coend}\\
&&\Set{x /| y}
\end{eqnarray*}

The basic noninterference requirement is that for every assertion $\Set{p}$ appearing in a thread, and for every update
$\Set{q} op$ in a concurrent thread, we have to prove $\Set{p /| q} op \Set{p}$. It's easily checked that the annotation 
above satisfies this condition. Finally, we require that the assertion at the beginning of each thread is implied by the
assertion immediately preceding the $\mathbf{cobegin}$, and at the conjunction of the assertions at the end of the threads
imply the assertion following the $\mathbf{coend}$.

Note that the noninterference rule prevents us from strengthening the assertion following the first $\mathbf{wait}$ to 
$x /| ~y$, because this assertion would not survive interference from the first assignment from the other thread.
(Similarly for the second thread, with the roles of $x$ and $y$ exchanged.) In fact, assuming there are no other
variables available, the annotation above is the strongest one that can be put on this program. Moreover, 
the postcondition happens to be valid when this program is run under weaker memory models like TSO (if
we flush out the store buffers at the end). 
Thus, we seem to have a potentially useful weak memory model that we can also reason about assertionally.

OG cognoscenti might wonder what's going on here, since the OG method is not only sound but complete for SCM.
But this completeness result required the unfettered use of {\emph{ghost code}}
 - data and code added to a program to make it easier to
reason about. (To make sure that adding ghost data and code is sound, ghost code must terminate, and is not allowed to write
to the non-ghost state.) So the difference between reasoning in our hypothetical weaker memory model and reasoning in SCM 
lies in the use of ghost code. In fact, we will see later that ghost updates can be added soundly to our weak
memory model, as long as doing so doesn't create new conflicts between writes of different threads.

Sadly, even without ghost code, our reasoning is not quite sound for all programs running under TSO. 
For example\footnote{
	This example is based on an example from \cite{x86TSO}, which was pointed out to the author by Peter Sewell.
}, consider the following:
\begin{eqnarray*}
&&\mathbf{cobegin}\\
&&\ \ \ \ \; x := \True;   \Wait{~y}; \\
&&\ ||\ \ y := \True;  \ x,b := \False,\True; \ \Wait{x}; \\
&&\mathbf{coend}\\
\end{eqnarray*}

We can prove the postcondition $\False$ as follows:
\begin{eqnarray*}
&&\mathbf{cobegin}\\
&&\ \ \ \ \;x := \True;  \Set{x \/ (b /| y)}\ \Wait{~y}; \Set{b<=>~x}\\
&&\ || \ \ y := \True;  \Set{y}\  x,b := \False,\True; \ \Set{b}\  \Wait{x}; \Set{b /| x}\\
&&\mathbf{coend}\\
&&\Set{\False}\\
\end{eqnarray*}

Nevertheless, this program has the following complete execution under TSO
 (where $\SB{s}$ means that the update $s$ emerges from the store buffer and hits memory):
\begin{eqnarray*}
&&x:=\True; \Wait{~y}; y:=\True; \SB{y:=\True};  x,b:=\False,\True; \SB{x,b:=\False,\True} ; \\
&&\SB{x:=\True}; \Wait{x}
\end{eqnarray*}

It turns out that the problem arises  because both threads race to write to the same variable, 
and that we can restore soundness for TSO by requiring a flush of the store buffer
whenever we write to such a variable. Note that this discipline does not require any flushes for the previous example,
where writes were racing only with reads,
whereas a discipline ensuring SCM does. So even though our weak memory model does not quite match TSO,
we can get to it from TSO much more efficiently than we can get to SCM. 

The main contribution of this paper is to define \Def{coherent causal} (CC) memory (CCM), a weak memory model in which
the kind of reasoning shown above is valid. CCM can be understood in several ways. First, it can be (roughly) understood as strengthening 
causal memory \cite{CausalMemory} by requiring that an execution must (causally) order conflicting (i.e. non-commuting) writes. 
This means that modulo the ordering of commuting writes, all threads see writes in the same global order; this is typically required
just to make sure that every thread sees the same final global ``state'' at the end of the day. 
However, a write need not be ordered with respect to conflicting reads (though it can be). 
(This distinction between reads and writes is essential in the example above; 
if the read of $y$ in the first thread above had to be causally ordered before the write to $y$ in the second thread, we would immediately
have a causality cycle, making a complete execution impossible.) 

A second view of CCM is as SCM extended with "weak reads" characteristic of causal memory
(like the $\Wait{}$ operations above), which only have to be valid in the context of the thread that performs them,
not in the the order observed by other threads. 
If we were to extend the state of each thread with a ghost variable recording its program counter, 
the weakness of weak reads would be manifest in the fact that we do not get to use the read itself when proving that 
the accompanying program counter update does not interfere with other threads. 
 
We show that Owicki-Gries reasoning is sound for CCM. 
Thus, it is also sound for any program running on causal memory
in which there are no ``write races'', i.e. executions where two threads concurrently write to the
same variable. This is a considerable generalization of the class of ``data-race-free'' programs (for causal memory) defined in \cite{CausalMemory} (which does not allow races as in the examples above), for which causal memory provides SC. 
This yields an assertional reasoning technique for a broad class of programs running under causal memory.

While CCM could be directly implemented by a hardware or software platform, we view CC primarily as a potential 
replacement for SC as a methodological target: instead of using a synchronization discipline (that guarantees
SC, one can use a more relaxed discipline that guarantees only CC. We give an example of such a discipline
for CC that guarantees that the program continues to simulate CC when run under TSO. This discipline is considerably
more relaxed than the one required for SC; it requires flushes only for writes that might participate in write races.
In particular, for any program in which no variable is written by more than
one thread, TSO already provides CC. 

\section{Coherent Causal Memory}

Fix a state space $S$. An \Def{operation} (notation: $o$) is a structure including
components $o.g$ (the \Def{guard} of $o$) and $o.u$ (the \Def{update} of $o$), 
where $o.g$ is a unary relation on $S$ and $o.g$ is a binary relation on $S$. 
In examples, we write operations as $\Op{o.g}{o.u}$ where $o.g$ is written
as a state predicate and $o.u$ is written as a command; if the guard is omitted,
it is by default $S$; if the update is omitted, it is by default the identity relation
on $S$.

A \Def{program} $P$ is a finite, partially ordered set of operations over $S$. 
We write $P(o)$ to mean that $o$ is an operation of $P$,
and $P(o_1,o_2)$ to mean that $P$ orders $o_1$ strictly before $o_2$. Define $P' \leq P$ ($P'$ is a \Def{prefix}
of $P$) by
\begin{eqnarray*}
P' \leq P &<=>&(A: o: P'(o) => P(o) /| (A: o': P(o',o) <=> P'(o',o)))
\end{eqnarray*}
If $P(o)$, define 
$\B:P:o$ ($P$ before $o$), $\NA:P:o$ ($P$ not after $o$), and $\Rem{P}{o}$ ($P$ without $o$)
to be the prefixes of $P$ satisfying
\begin{eqnarray*}
\B:P:o(o') &<=>& P(o',o)\\
\NA:P:o(o') &<=>& P(o') /| ~P(o,o')\\
(\Rem{P}{o})(o') &<=>& \NA:P:o(o') /| o \neq o'\\
\end{eqnarray*}

An \Def{execution} $E$ of $P$ is
a program with the same operations
as $P$ (but possibly a stronger ordering relationship on the operations), 
along with a map $\Red{}$ from prefixes of $E$ to $S$, satisfying 
\begin{eqnarray}
\label{E1}&&(A: o: E(o) => o.g(\Red{\B:E:o}))\\
\label{E2}&&(A: o, E': E' \leq E /| E' = \NA:E':o => o.u(\Red{\Rem{E'}{o}}, \Red{E'})
\end{eqnarray}
We can think of the prefixes of $E$ as the ``states'' of the execution that
might be seen by various observers; $\B:E:o$ is the state seen by operation $o$. 
\Ref{E1} says that the guard of each operation holds in the state
that it sees; \Ref{E2} says that if $o$ is a terminal operation of a state $E'$,
then the state can be obtained by applying the update $o.u$ to the ``preceding''
state with $o$ removed. 

The guard $o.g$ can be viewed as a generalized read. For example, an operation that
copies the value of a state variable $v$ to state variable $r$ can be represented as
the operation $\Op{v=c}{r:=c}$ where $c$ is a fresh state constant (i.e., a state 
variable that is never updated). In any execution,
$c$ gives the value read from $v$ in this operation; since $c$ is fresh, an
implementation could choose it ``lazily'' (as the value of $v$ in
the state where the operation executes).

Note that if the updates of all pairs of operations unordered in $E$ commute, then 
$\Red{}$ is uniquely determined from the initial state,
so $E$ is an execution iff each guard holds in the corresponding state. 

CC can be viewed as an extension of SC with the addition of guards;
if we ignore guards (e.g., make them all $\True$), CC is
essentially equivalent to SC, because 
every execution can be extended to a linear execution with the same final state.
This does not hold for programs in general, because extending an execution to a linear
order changes the prestates of operations, possibly resulting in states that
violate their guards. For example, the following program, where 
the precedence relation is given by the transitive closure of the
arrows shown, has a CC execution starting from any initial state, but
no SC executions:

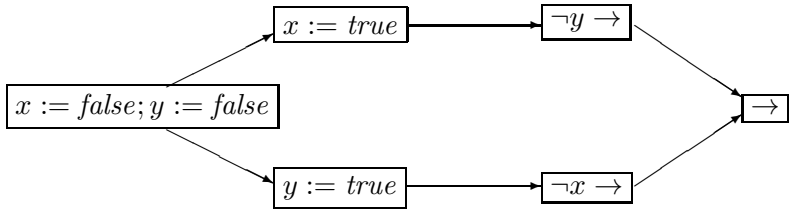
\def\DOp#1#2{\framebox{$#1$}}
\noMath
\setlength{\unitlength}{0.14in}
\def\Example{
\put(0,3){\DOp{x := \False; y := \False}}
\put(10,6){\DOp{x := \True}}
\put(6,4){\vector(2,1){4}}
\put(20,6.3){\DOp{\Op{\neg y}{\Skip}}}
\put(15,6.3){\vector(1,0){5}}
\put(10,0){\DOp{y := \True}}
\put(6,2.4){\vector(2,-1){4}}
\put(20,0){\DOp{\Op{\neg x}{\Skip}}}
\put(15,0.3){\vector(1,0){5}}
\put(27.5,3){\DOp{\Op{}{}}}
\put(23.5,0.3){\vector(3,2){4}}
\put(23.5,6.3){\vector(3,-2){4}}
}

\begin{picture}(32,10)(0,-1)
\Example
\end{picture}
\Math

\section{Annotations}

Let $P$ be a program, and let $A$ be a map from ordered pairs of
$P$ operations to unary predicates on $S$. Define 
$A(o)(s) <=> (A: o': P(o',o) => A(o',o)(s))$. $A$ is an \Def{annotation} of $P$ iff the following
conditions hold: 
\begin{eqnarray}
\label{A3} &&(A: o,o',s,s': o.g(s) /| A(o)(s) /| o.u(s,s') => A(o,o')(s'))\\
\nonumber &&(A: o,o',o'',s,s': ~P(o'',o) /| o''\neq o /| ~P(o',o'') /|  o' \neq o'' \\
\label{A4}&& \hspace{1in} /| A(o,o')(s) /| A(o'')(s) /| o''.u(s,s') => A(o,o')(s'))
\end{eqnarray}

Intuitively, $A(o,o')$ should hold in any state where $o$ has been executed but $o'$ has not,
and $A(o)$ should hold in the state seen by $o$. The requirements say that \Ref{A3}
executing $o.u$ in any state satisfying $o.g$ and $A(o)$ 
results in a state satisfying each annotation going out
from $o$ (``local correctness''); and \Ref{A4} the update of any operation $o''$
that can occur between $o$ and $o'$ must preserve the annotation on
the edge from $o$ to $o'$ (``noninterference'').
Note that for a program written as a union of linear orders,
with all guards $\True$ and nontrivial annotations only between successive operations,
an annotation is just an OG proof outline.

Here is our example program, given the strongest annotation possible: 

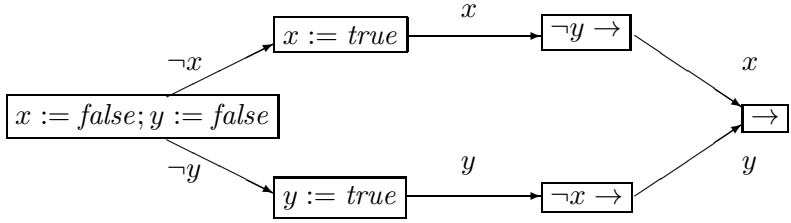
\def\Annotation#1{\makebox{$#1$}}
\noMath
\begin{picture}(32,10)(0,-1)
\Example
\put(6,5){\Annotation{\neg x}}
\put(17,7){\Annotation{x}}
\put(27.5,5){\Annotation{x}}
\put(6,1){\Annotation{\neg y}}
\put(17,1.3){\Annotation{y}}
\put(27.5,1.3){\Annotation{y}}
\end{picture}
\Math

\section{Soundness}
Let $A$ be an annotation of program $P$. We prove that if $E$ is an execution of a prefix of $P$, 
$E(o)$, and $~E(o_1)$, then $A(o,o_1)(\Red{E}))$. 

The proof is by induction on the size of $E$. 
If $E$ has an operation $o_2$ s.t. $E = \NA:E:{o_2} /| o_2 \neq o$, then 
\begin{Proof}
A(o,o_1)(\Red{E}) &<=& E_1 := \Rem{E}{o_2};\ \Ref{A4}\\
A(o,o_1)(\Red{E_1}) /| A(o_2)(\Red{E_1}) /| o_2.u(\Red{E_1},\Red{E})) &<=&\Ref{E2}\\
A(o,o_1)(\Red{E_1}) /| A(o_2)(\Red{E_1}) &<=&\Induc\\
A(o_2)(\Red{E_1})&<=&\mbox{def of }A(o_2)\\
(A: o_3: P(o_3,o_2) => A(o_3,o_2)(\Red{E_1})) &<=&\Induc\\
\True
\end{Proof}

Conversely, if there is no such $o_2$, then $E = \NA:E:o$, and $\Rem{E}{o} = \B:E:o$, so
\begin{Proof}
A(o,o')(\Red{E}) & <= &\Ref{A3}; \Rem{E}{o} = \B:E:o\\
o.g(\Red{\B:E:o}) /| A(o)(\Red{\B:E:o}) /| o.u(\Red{\B:E:o}, \Red{E}) &<=& \Ref{E2}\\
o.g(\Red{\B:E:o}) /| A(o)(\Red{\B:E:o}) &<=& \Ref{E1}\\
A(o)(\Red{\B:E:o}) &<=& \mbox{def of }A(o)\\
(A: o'': P(o'',o) => A(o'',o)(\Red{\B:E:o})) &<=&\Induc \\
\True 
\end{Proof}

\section{Ghosts}
Annotations are not complete, even for programs with only trivial guards,
for essentially the same reason that
they are not complete in the original OG theory: we need
some way to express assertions about the ``program counters'' of other
``threads''. As usual, we achieve completeness by allowing ghost updates.
However, we must be careful; unrestricted use of ghost updates would make the proof
system  complete for
SC, and hence is unsound for CC.

It is instructive to see where the problem arises for our example program. 
Suppose we wanted to prove the precondition $\False$ for the final operation,
i.e. that the program has no feasible executions. Using OG, we could do this by
adding a ghost variables to record which thread ``won the race'' to 
read the flag set by the other thread. One way would be to
introduce two Boolean ghost variables $xw$ (``$x$ won'') and $yw$
(``$y$ won''), decorating and annotation the program as follows, and giving
each edge an additional conjunct of the global invariant
\[ (xw => x /| ~yw) /| (yw => y /| ~xw) \]

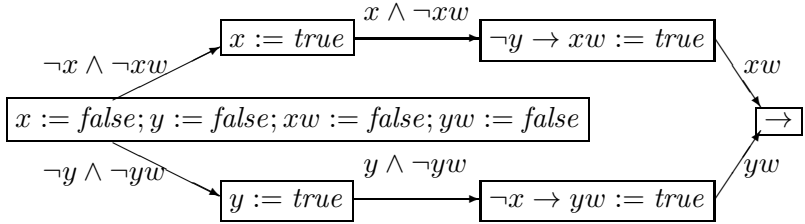
\def\Annotation#1{\makebox{ $#1$ }}
\noMath
\begin{picture}(32,10)(0,-1)
\put(0,3){\DOp{x := \False; y := \False; xw := \False; yw := \False}}
\put(8,6){\DOp{x := \True}}
\put(4,4){\vector(2,1){4}}
\put(17.7,6){\DOp{\Op{\neg y}{xw :=\True}}}
\put(13.0,6.3){\vector(1,0){4.7}}
\put(8,0){\DOp{y := \True}}
\put(4,2.4){\vector(2,-1){4}}
\put(17.7,0){\DOp{\Op{\neg x}{yw := \True}}}
\put(13,0.3){\vector(1,0){4.7}}
\put(28,3){\DOp{\Op{}{}}}
\put(26.5,0.3){\vector(2,3){1.7}}
\put(26.5,6.3){\vector(2,-3){1.7}}

\put(1,5){\Annotation{\neg x \AndSym \neg xw}}
\put(13,7){\Annotation{x \AndSym \neg xw}}
\put(27.2,5){\Annotation{xw }}
\put(1,1){\Annotation{\neg y \AndSym \neg yw}}
\put(13,1.3){\Annotation{y \AndSym \neg yw}}
\put(27.2,1.3){\Annotation{yw }}
\end{picture}
\Math

For the $\Skip$ operation, the global invariant, conjoined with $xw$
and $yw$, gives the desired annotation $\False$.
However, this annotation is unsound for CC, because the annotation on
the edge labeled $yw$ (which implicitly includes the global 
invariant $xw =>~yw$)  is not preserved by the update $xw := \True$, even
assuming the  precondition of this update ($x /| ~xw$), because
we cannot use the guard of the update in proving noninterference.
Note that this annotation would be sound if we replace the ``weak
read'' $\Op{~y}{xw := \True}$ with a ``strong read'' 
$\Assume{~y}; xw := \True$ (and similarly with the update to $yw$);
however, these strong reads conflict with the corresponding writes to
$y$ and $x$ respectively, and so would have to be ordered before these
actions in any execution (which immediately implies there are none).

Alternatively, we could introduce a ghost update to keep track of
the race between the assignments to $x$ and $y$ (with the same
global invariant as before):

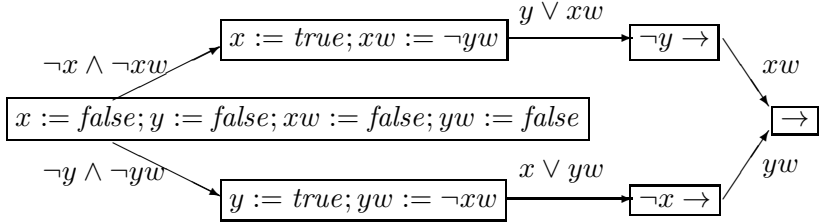
\def\Annotation#1{\makebox{ $#1$ }}
\noMath
\begin{picture}(32,10)(0,-1)
\put(0,3){\DOp{x := \False; y := \False; xw := \False; yw := \False}}
\put(8,6){\DOp{x := \True; xw :=\neg yw}}
\put(4,4){\vector(2,1){4}}
\put(23.3,6){\DOp{\Op{\neg y}{\Skip}}}
\put(18.7,6.3){\vector(1,0){4.7}}
\put(8,0){\DOp{y := \True; yw := \neg xw}}
\put(4,2.4){\vector(2,-1){4}}
\put(23.3,0){\DOp{\Op{\neg x}{\Skip}}}
\put(18.7,0.3){\vector(1,0){4.7}}
\put(28.6,3){\DOp{\Op{}{}}}
\put(26.8,0.3){\vector(2,3){1.7}}
\put(26.8,6.3){\vector(2,-3){1.7}}

\put(1,5){\Annotation{\neg x \AndSym \neg xw}}
\put(18.8,7){\Annotation{y \OrSym xw}}
\put(27.9,5){\Annotation{xw  }}
\put(1,1){\Annotation{\neg y \AndSym \neg yw}}
\put(18.8,1.1){\Annotation{x \OrSym yw}}
\put(27.9,1.3){\Annotation{yw}}
\end{picture}
\Math

This annotation satisfies the noninterference condition. However, 
the augmented program (with the 
ghost updates) does not simulate the original program: there are executions 
of the original program that do not correspond to executions 
of the augmented program because of the
coherence condition \Ref{E2}. In essence, the ghost updates have
introduced a race between the updates to $x$ and $y$, which were not racing in
the original program; the race forces executions of the augmented program
to order these updates, which might have remained unordered in an execution
of the original 
program. 

One condition on ghost updates that suffices to make them sound is the following.
If there is a state $s$ from which two updates from $P$ can coherently execute in parallel,
then for any state with concrete state $s$, there is a coherent execution of the augmented 
updates that projects to the same coherent execution. In the case where the updates are
partial functions, we can restate this as follows: 
if two updates from $P$ commute from a state $s$, then
the corresponding augmented updates also commute from states that project to $s$. 

Given a program $P$, we can introduce for each operation $o$ a Boolean ghost variable
$o.done$ ($\True$ iff $o$ has executed) and a ghost constant
$o.in$ which gives the concrete state seen by operation $o$. (That is, we implicitly strengthen
$o.g$ to $o.g /| \langle \mbox{concrete state} \rangle =  o.in$.) We add to $o$ the ghost update
$o.done := \True$; this obviously satisfies the soundness condition above on ghost updates. 
Finally, we can define $A(o,o')$ as 
$(E:E: o.done /| ~o'.done /| (A:o'': o''.done <=> E(o'')) /| (A:o'': E(o'') => \B:E:{o''} = o''.in) /|
\langle \mbox{concrete state}\rangle = \Red{E})$
where $E$ ranges over executions of prefixes $P$. This annotation gives completeness,
i.e., $A(o,o')(s)$ iff there exists an execution $E$ where $s$ is the concrete state of $E$.

The limitations on ghost updates likewise suggests a new programming language feature: if we want to
be able to add ghost updates that create conflict between actions whose concrete updates do not conflict,
we should be able to introduce such ``virtual conflict'' declaratively, i.e. force the compiler/runtime
to order the concrete updates. Since it is programmatically 
inconvenient to talk about actions in other threads per se, a natural way to do this is via ``conflict variables'',
which can be viewed as resources that acquired and released as part of an atomic update. Two updates that
conflict on such variables are required to be ordered with respect to each other. The use of conflict variables means
that we have complete freedom in the use of ghost code, but that we may have to add updates to conflict variables 
to some updates that simultaneously update ghost state. This gives us a fine-grained way to add to a program
only as much synchronization as is necessary to justify the property we are trying to prove, giving an ideal blend
of (formal) programmability and implementation efficiency.

\section{A CC discipline for TSO}

For illustration, we describe a sound synchronization discipline to implement CC on top of TSO. It is not the most general 
such discipline, or even a generally useful one. We include it only to illustrate the advantage of reducing to CC instead of to SC.

Assume that our program $P$ consists of a disjoint union of linear orders, each of which we call a \Def{thread}. 
Each memory location is either \Def{shared} or \Def{unshared}; each unshared variable has a unique (thread) \Def{owner}.
Each operation of each thread is either a read of a variable $v$ (i.e., an operation of the form $\Op{v=val}{}$, where $val$
is a constant), a write of a variable, or a read-modify-write of a variable (of the form $\Op{v=val}{v := val'}$);
in the latter two cases, the variable must either be shared or owned by the thread.

The program executes as follows.  A \Def{buffered write} is a pair consisting of a 
variable name and a value. A store buffer is a sequence (i.e., queue) of buffered writes. Associated with each thread 
$t$ is a store buffer $b[t]$, which is
initially empty. It is an invariant of the following operational model that every buffered write in $b[t]$ is to a variable owned by $t$.

The actions of the system are as follows. A write by thread $t$ of value $val$ to a location $v$ owned by $t$ is executed by an ordinary
(buffered) write (which appends the write to $b[t]$). A write by $t$ to a shared location, or a read-modify-write
of a location, is implemented by an interlocked operation, which can only execute when $b[t]$ is empty, and leaves $b[t]$
empty, performing the update directly on the memory as an atomic action. A read by $t$ of variable $v$ 
returns the most recently buffered write to $v$ in $b[t]$ (if such a write exists), and otherwise returns $m[v]$.
This operation blocks if the read value does not satisfy the guard of the read operation.
Finally, if $b[t]$ is nonempty, the buffered write at the front of the store buffer can be removed from the buffer and applied to the memory.

Corresponding to each such execution on TSO that completes, we can define an order on operations as follows. 
For operations $o,o'$,  $E(o',o)$ iff $P(o',o)$ or if there exists a write $o''$ that completes to memory before $o$ executes 
and ($o'=o'' \/ P(o',o''))$. (This is transitive by definition.) 
Note that the only conflicting operation updates are writes or read-modify-writes to a shared locations by different threads,
which by the definition above are ordered (one way or the other), because such actions are implemented atomically
(so one action will have hit memory before the other executes).
Thus, $E$ defines a CC execution iff each guard is satisfied in the
state determined by its prefix. By the definition above, the guard $v=val$ holds in the prestate of an operation iff
an ordinary TSO read of $v$ returns $val$. Thus, the generated execution is a complete TSO execution iff the derived order
is a CC execution. 
 
\section{Related Work}
Many theoretical models of concurrent and distributed systems have been proposed. For example, there
are models that eliminate state entirely and define a program or an execution as a partially ordered sets of events, e.g. \cite{Winskel},
defining states as cuts (i.e., sets of events closed under predecessor). As far as we know, none of these approaches
have led to a usable approach for reasoning about concurrent software.

A number of theorems that give conditions under which a program executing on a given memory model
simulates SCM. Some (e.g. \cite{WeakOrdering}) have gone so far as to propose that memory models should  
be defined in terms of a programming discipline sufficient to make programs running on the model SC.
Most useful memory models provide SC for data-race-free programs, 
and more sophisticated reduction theorems exist for more general classes of programs
running on particular memory models. For example, \cite{SBReduction} gives a programming discipline
for TSO; this discipline is expressed as a set of invariants that must be met by the program when running under SCM,
allowing all program reasoning to be carried out assuming SCM. This is particularly useful when the conditions are being
discharged using a standard program verification system.
An alternative (though less powerful) way to define a suitable condition is through a type system; for example, 
a program well-typed under concurrent separation logic \cite{CSL} is sequentially consistent when run under TSO
if each conditional critical region can be implemented with an atomic memory action that flushes the store buffer.

\bibliography{bibliography}

\end{document}